# Turning Cluster Management into Data Management: A System Overview

Eric Robinson
University of Wisconsin - Madison
1210 W. Dayton St.
Madison, WI 53706
erobinso@cs.wisc.edu

David J. DeWitt
University of Wisconsin - Madison
1210 W. Dayton St.
Madison, WI 53706
dewitt@cs.wisc.edu

## ABSTRACT
This paper introduces the CondorJ2 cluster management system. Traditionally, cluster management systems such as Condor employ a process-oriented approach with little or no use of modern database system technology. In contrast, CondorJ2 employs a data-centric, 3-tier web-application architecture for all system functions (e.g., job submission, monitoring and scheduling; node configuration, monitoring and management, etc.) except for job execution. Employing a data-oriented approach allows the core challenge (i.e., managing and coordinating a large set of distributed computing resources) to be transformed from a relatively low-level systems problem into a more abstract, higher-level data management problem. Preliminary results suggest that CondorJ2's use of standard 3-tier software represents a significant step forward to the design and implementation of large clusters (1,000 to 10,000 nodes).

## 1. INTRODUCTION
At one end of the systems spectrum is the traditional batch computing system that employs a process-centric design with data accessibility and location-independent access largely an afterthought. Such systems frequently generate vast amounts of data that is an indispensable source of information for both users and system administrators. Unfortunately, efficiently accessing and manipulating this data is often difficult or impossible. From a system design and development perspective, these process-centric architectures often lead to extensibility (e.g., development-time and run-time overheads required to add additional system services and/or functionality), scalability (e.g., the number of compute nodes and/or jobs the system can manage) and performance (e.g., the ease and speed with which users and administrators can extract actionable data from the system) barriers that are difficult to overcome.

At the opposite end of the spectrum is the n-tier web-application in which data accessibility and location-independent access are top priorities and OS-level process management is essentially unheard of. The data-centric nature of these systems means that the most difficult task in extending the system is often simply getting the underlying data-model right; with a proper model in place implementing the application logic often follows naturally. Additionally, these systems employ a variety of techniques specifically designed to overcome scalability and performance barriers.

An interesting research question is whether it is possible to combine the strengths of these two complementary approaches in order to build a scalable, high-performance system that provides users and administrators with location-independent access to a large collection of distributed computing resources. This paper asserts that the answer to this question is yes, and introduces the CondorJ2 cluster management system as evidence. We assert that it is advantageous to adopt a process-centric approach for the distributed system components but a data-centric approach for the centralized system components and glue the two together with a standardized messaging protocol. Viewed in this light, overcoming limitations to extensibility, scalability or performance is reduced to a data management problem. The extensibility of the system architecture, for example, is largely driven by the extensibility of the underlying data model. Similarly, system performance and scalability rests on the speed and efficiency with which incoming messages can be transformed into actions on the underlying database.

## 2. OVERVIEW OF A PROCESS-CENTRIC SYSTEM
The purpose of this section is to provide a general overview of the design pattern of the dominant batch computing systems including Condor [5], LSF [9], PBS [8], IBM's LoadLeveler [4] and Sun's N1 GridEngine [1], among others. While each of these systems has a unique design and feature set, the underlying architectures are quite similar. Thus, for brevity, this section only describes Condor [2,5].

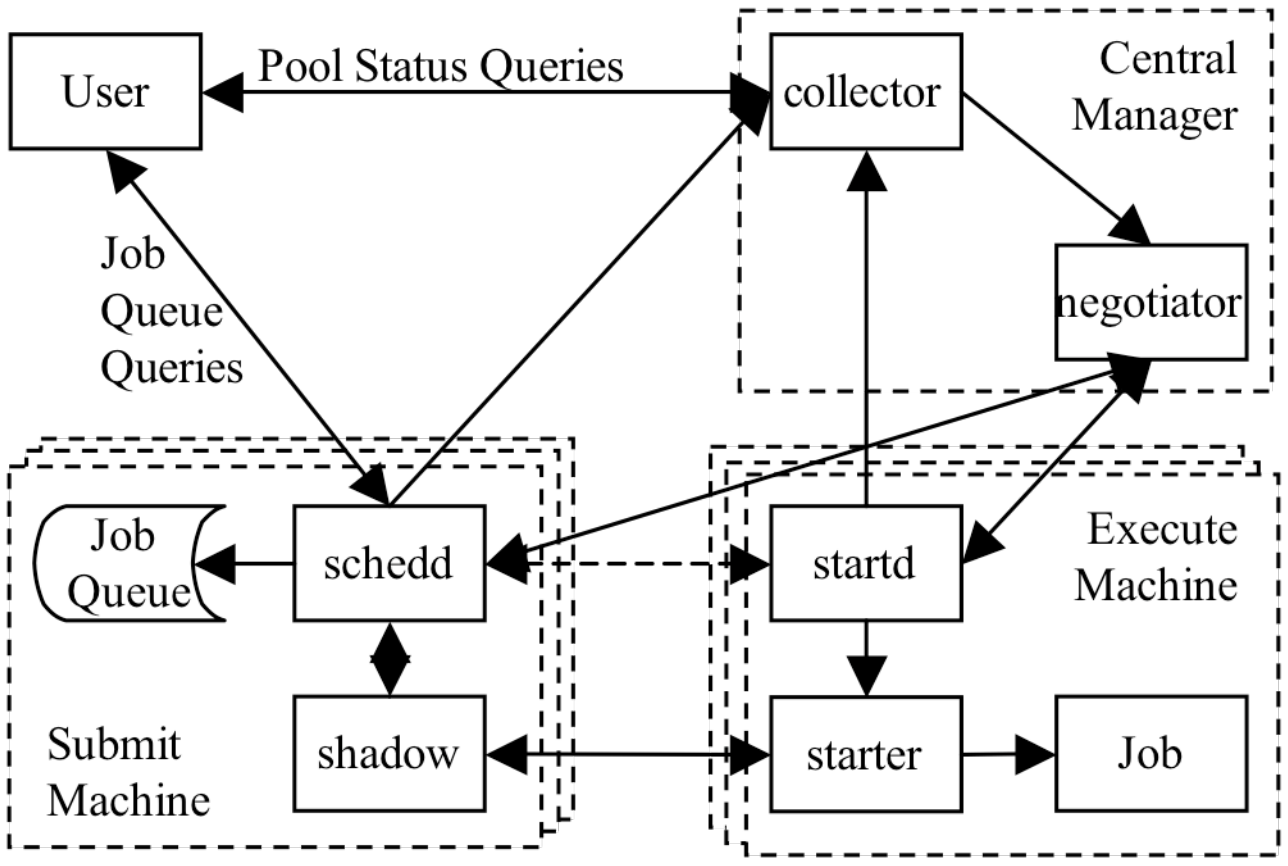

**Figure 1. Condor Architecture**

Figure 1 shows the architecture of Condor, which employs a semi-distributed, process-oriented architecture. Conceptually, Condor provides three basic services: job submission and monitoring, matchmaking, and job execution. The three services are mutually compatible and a machine can be configured to provide one, two, or all three of these services. Matchmaking (the task of assigning jobs to machines) for a Condor pool is a centralized process so for a given pool only a single machine performs matchmaking. Submission and execution services, however, can be provided by any number of machines in the pool. Six daemons, two for each service type, work cooperatively to shepherd jobs through the system until completion. A seventh daemon, the master, runs on every machine in the pool. The master daemon is responsible for monitoring the other daemons and restarting a daemon if it fails.

## 2.1 Overview of Job Submission

Users can submit jobs to a Condor pool from a submit machine. Any machine in the pool can act as a submit machine. The daemons that run on a submit machine are the *schedd* (pronounced "sked-dee") and the *shadow*.

The *schedd* serves as the job-queue manager for the machine that it is running on. When users submit jobs to a Condor pool they are actually submitting jobs to the *schedd* that is running on a particular machine. The *schedd* uses persistent storage (an OS file) and transactional semantics to guarantee that no submitted jobs are lost and to ensure appropriate behavior upon recovery from process or machine failure. In addition to handling job submissions, the *schedd* is also responsible for responding to users' job-queue related queries. It is important to note that the persistent version of the job queue is maintained only for fulfilling the transaction and recovery guarantees outlined above. For operational purposes, such as responding to user- or system-generated queries, the *schedd* relies on an in-memory version of the queue. Since the schedd is a single-threaded process it needs no concurrency logic for job queue management functions.

Once a job in the queue has been matched to a machine to run on, the *schedd* spawns a *shadow*. The *shadow* is responsible for monitoring the remote execution of the job and responding to resource requests from the job [6]. Additionally, for jobs that use Condor's file transfer mechanism it is the *shadow* that sends the input file(s) to the execute machine and receives the output and error files from the execute machine. Note that, in contrast to the *schedd*, the shadow implements no transaction or recovery logic. Also note that the one-to-one relationship between a *shadow* and an executing job means that, at a particular point in time, a given submit machine will have a *shadow* process running for every currently executing job submitted from that machine.

## 2.2 Overview of Matchmaking

Matchmaking in Condor is a centralized process in which machine and job information is examined in order to match jobs to machines in a way that reflects the various pool, user, machine, and job priority policies [10]. The matchmaking process is comprised of two steps – information collection and resource allocation. Reflecting this, two daemons, the *collector* and the *negotiator*, collaborate to implement matchmaking in Condor. Despite the fact that the *collector* and *negotiator* could technically run on separate machines, matchmaking is still a truly centralized process because there can only be one *collector* and one *negotiator* for a given pool at a given time.

The *collector* daemon serves as a central repository for machine and job information in a given pool. Submit machines and execute machines periodically send status updates to the *collector*. The *collector* maintains all of this information in memory. The *collector* is responsible for forwarding this information to the *negotiator* upon request. The *collector* is also responsible for responding to pool-level queries. Since job queues are managed by individual *schedd*s and since machine and job queue information is periodically refreshed, the *collector* needs no transaction or recovery logic. Upon restart after a failure the *collector* rebuilds its in-memory data structure as updates arrive.

The *negotiator* performs the matchmaking required to make job-scheduling decisions. To initiate a negotiation cycle, the *negotiator* queries the *collector* to obtain the necessary data. The *negotiator* is responsible for making resource allocation decisions subject to machine and job specific requirements and various priority policies. The *negotiator* maintains all data and performs all calculations in memory. Since each negotiation cycle is more or less independent, the *negotiator* needs no transaction or recovery logic. Note that negotiation can only occur if both the *collector* and the *negotiator* are functioning properly. If either the *collector* or the *negotiator* were to fail, users could continue to submit jobs (and any currently executing jobs would continue to run) but no new matches could be made. Once both the *collector* and the *negotiator* are back online matchmaking begins again.

## 2.3 Overview of Job Execution

Jobs in a Condor pool run on an execute machine. As with a submit machine, any machine in the pool can act as an execute machine and there can be any number of execute machines in a given pool. The daemons that run on an execute machine are the *startd* and the *starter*.

The *startd* serves as the representative for the machine that it is running on. It is the *startd*'s responsibility to gather the relevant data about its machine and to periodically send this data to the *collector*. It is also the *startd*'s responsibility to monitor the activity on the machine and to enforce the policy that the machine owner has defined for when and how that machine is to be made available to the Condor pool. Once an execute machine has been assigned a job to run, the *startd* on that execute machine will spawn a *starter* daemon to set up the actual execution of the job.

The *starter* daemon is responsible for setting up the execution environment and spawning the actual job that is to run. The *starter* monitors the job as it executes and notifies the associated *shadow* when important job events (e.g., startup, completion, etc.) occur. Additionally, the *starter* communicates with the corresponding *shadow* running on the job's submit machine to implement the Condor file transfer mechanism (if necessary) noted above. Neither the *startd* nor the *starter* implements any transaction or recovery logic.

# 3. OVERVIEW OF DATA-CENTRIC WEB APPLICATIONS

This section provides a brief overview of a typical three-tier architecture used by many enterprise web applications. RDBMSs and HTTP Servers are key components of this architecture. We assume that readers are already familiar with their basic functionality, so the focus here is on the third component of the web application infrastructure – the application server.

Application servers provide a broad range of reusable functionality that simplifies development activities by abstracting away low-level intricacies of commonly required services such as database connection pooling, the two-phase commit protocol, client authentication, etc. The application server market is currently split between Microsoft's .NET platform and the Java 2 Enterprise Edition (J2EE) platform. The following subsections focus on the details of J2EE application servers because CondorJ2 was built on this platform.

## 3.1 J2EE and EJB Features

In this section we give a brief overview of the services that J2EE application servers provide. Figure 2 illustrates the structure of a basic J2EE + EJB (Enterprise Java Bean) application [11,3]. For brevity, a lexicon of J2EE and EJB terminology is omitted; this should not affect the reader's ability to distill the main points.

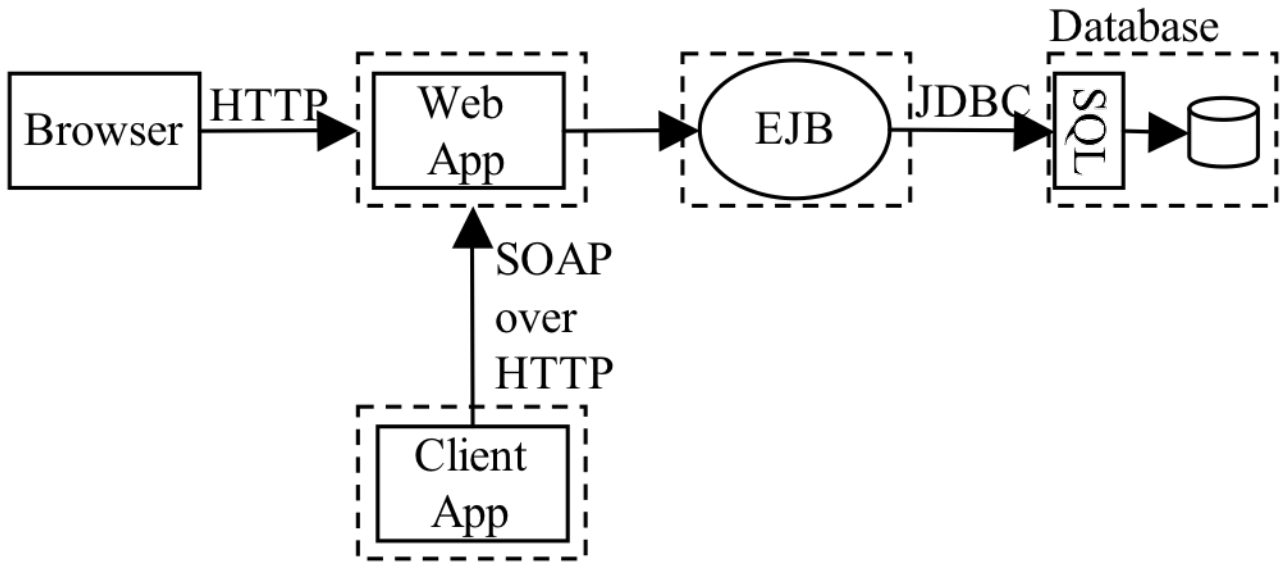

**Figure 2. J2EE Architecture (Adapted from [7])**

J2EE application components are deployed into containers. Containers provide a standard set of services, and a standardized API for accessing those services [11]. An application server product such as IBM's WebSphere, BEA's WebLogic, Oracle's Application Server 10g, JBoss AS, Sun's Sun ONE, etc. implements the containers into which J2EE and EJB applications are deployed. In Figure 2 above, the Web Container and EJB container would be running inside an Application Server (which also contains an embedded web server). The browser could be any Internet browser, while the web client would be any agent capable of constructing and sending a SOAP request over HTTP. The database would be any data storage application that provides a JDBC interface and is registered with the Application Server.

One of the key features of the J2EE + EJB architecture is the decoupling of application development tasks from application deployment tasks. This decoupling allows for a greater level of abstraction for J2EE application developers and a greater level of site-specific customization at application deployment time. Some additional, relevant key features of this architecture are [7]:

•**Portability:** Because the J2EE and EJB architecture has been standardized, applications that conform to the standard will run in any J2EE compliant Application Server.

•**Business logic independent from presentation logic:** The presentation portion of the application is handled in the Web container (html, jsp), while the business portion is handled in the EJB container.

•**Container Managed Persistence:** Developers can elect to allow the container to manage the persistence (in the database) of entity beans. Not only does this allow the container more flexibility for data access optimizations, but it also frees the developer from implementing (and debugging) the object to relational mapping.

•**Database connection pooling:** The container provides the necessary connection pooling mechanisms. This is especially powerful when used in combination with container-managed persistence for entity beans because it gives the container the opportunity to implement optimizations to reduce database connection contention even further.

•**Clustering:** Application servers (including the free, open-source JBoss AS product) provide the ability to "cluster" the application server across multiple physical machines. This allows for a greater number of concurrent clients as well as providing high-availability if some of the clustered machines should go down.

# 4. CONDORJ2

Figure 3 contains a diagram of the CondorJ2 architecture consisting of a central database and a J2EE + EJB application deployed in an application server.

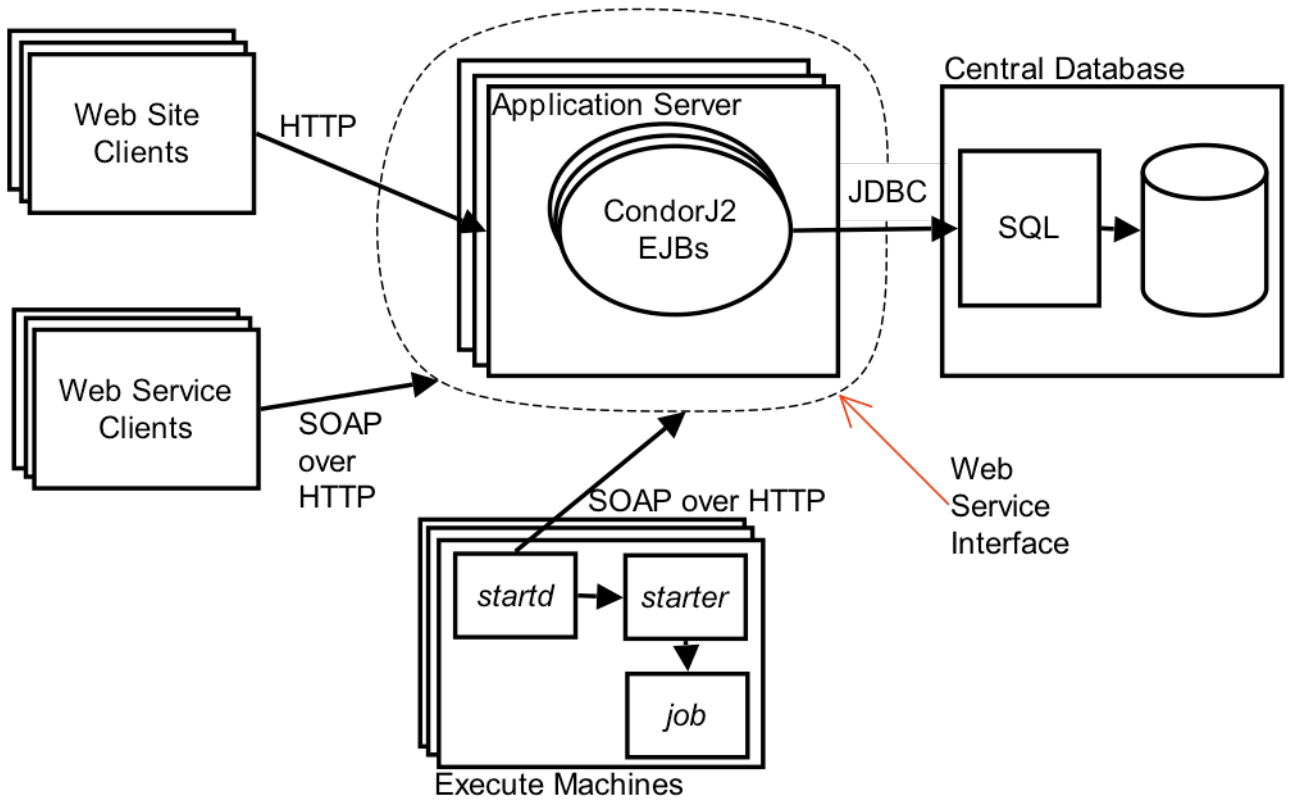

**Figure 3. CondorJ2 Architecture**

At a high level, CondorJ2 differs from Condor in three fundamental ways:

1) An RDBMS provides improved data accessibility, high concurrency, transaction and recovery services, and an expressive query language over the operational data.

2) A single system-wide job repository replaces the stand-alone submit machines allowing the computation management and job scheduling components to be tightly coupled. The communication and replication overheads eliminated as a consequence significantly improve both system scalability and performance. Additionally, the fact that these two components share common data structures eliminates the need to manage a shadow process in order to monitor running jobs.

3) An application server makes it possible to keep up with the message traffic for even very large pools while its connection-pooling and persistence management mechanisms reduce the required number of simultaneous open connections to the database. In addition, modeling the persistent system data using entity beans means that the relationships present among data items are explicitly declared. The developer thus has an intuitive means for rapidly navigating between pieces of related data.

## 4.1 Architecture Overview

The components of the CondorJ2 architecture can be divided into three classes – the CondorJ2 Application Server (or CAS), the database and the clients. Note that the CAS is the only entity in the system with direct access to the database. All non-CAS CondorJ2 processes that need to read or update the operational data interact with the CAS, which, in conjunction with the RDBMS, assumes responsibility for managing the necessary concurrency, querying and persistence mechanisms. Note also that the term "clients" is overloaded – it refers not only to the traditional notion of actual CondorJ2 pool users and administrators but also to all of the pool's execute machines. Here the notion of a client has been expanded to include the pool's execute machines to reflect the fact that, in terms of accessing and updating system data, both traditional clients and execute machines are clients of the CAS.

There are three distinct entities that need to be able to communicate with the CAS: users/administrators, web clients and execute machines. Users and administrators communicate with the CAS via a CondorJ2 pool's web site. Users and administrators submit jobs, access standard reports, pose queries and configure system behavior from anywhere that they have access to the web. For web clients the ability of the CAS to expose beans as web services provides a clear path for providing an interface to pool-services as a single standards-compliant set of web services. For daemons running on execute machines, the CAS exposes a set of web services specifically tailored to the interactions the daemons need to have with the operational data store. For the *startd*, for example, the CAS exposes the "beginExecute" web service that the *startd* invokes whenever it needs to alert the application server that it is going to begin executing a job. The bean behind the web service shepherds the data through the system and reports success or failure back to the *startd*.

The server-side CondorJ2 application-tier code is implemented via a layered approach. Figure 4 contains a diagram describing the implementation. The persistence layer consists of the entity beans that represent the persistent objects (e.g., users, workflows, jobs, machines, configuration policies, etc.) that collectively determine system state. There is a one-to-one correspondence between entity bean objects and tuples in the underlying database[1]. Additionally, each entity bean provides a well-defined service-oriented interface of operations that can be invoked on that particular object. These operations translate into SELECT, UPDATE, INSERT or DELETE operations on the tuples in the database. Embedded within the entity bean implementation is the application logic required to a) verify that the object is in a state in which the particular service call is valid, b) perform the requested operation and c) verify that the service invocation did not leave the object in an inconsistent state. Employing this approach system-wide yields a large set of fine-grained services. Directly exposing such fine-grained functionality to extra-system entities (i.e., clients) is a bad idea for two reasons. First, the granularity of service desired by a client is generally coarser than the granularity of service required to maximize architectural efficiency. The client should be empowered to interact with the system at the appropriate level, in ignorance of the underlying architecture. Second, requiring remote web clients to compose commonly required coarse-grained services from fine-grained system-level components results in an unacceptable level of communication overhead that could limit system performance and scalability.

This "granularity mismatch" is resolved in an application logic layer that wraps the persistence layer. This layer provides the client-appropriate coarse-grained services composed of the fine-grained entity bean services. All interaction with the system goes through this application logic layer; nothing besides the application logic layer communicates directly with the persistence layer. Built on top of the application logic layer are the two external interfaces to CondorJ2 - the pool web site and the web service interface. Because both external interfaces are built on top of the same set of underlying system services, they are capable of offering identical functionality – the only difference being the presentation to the client.

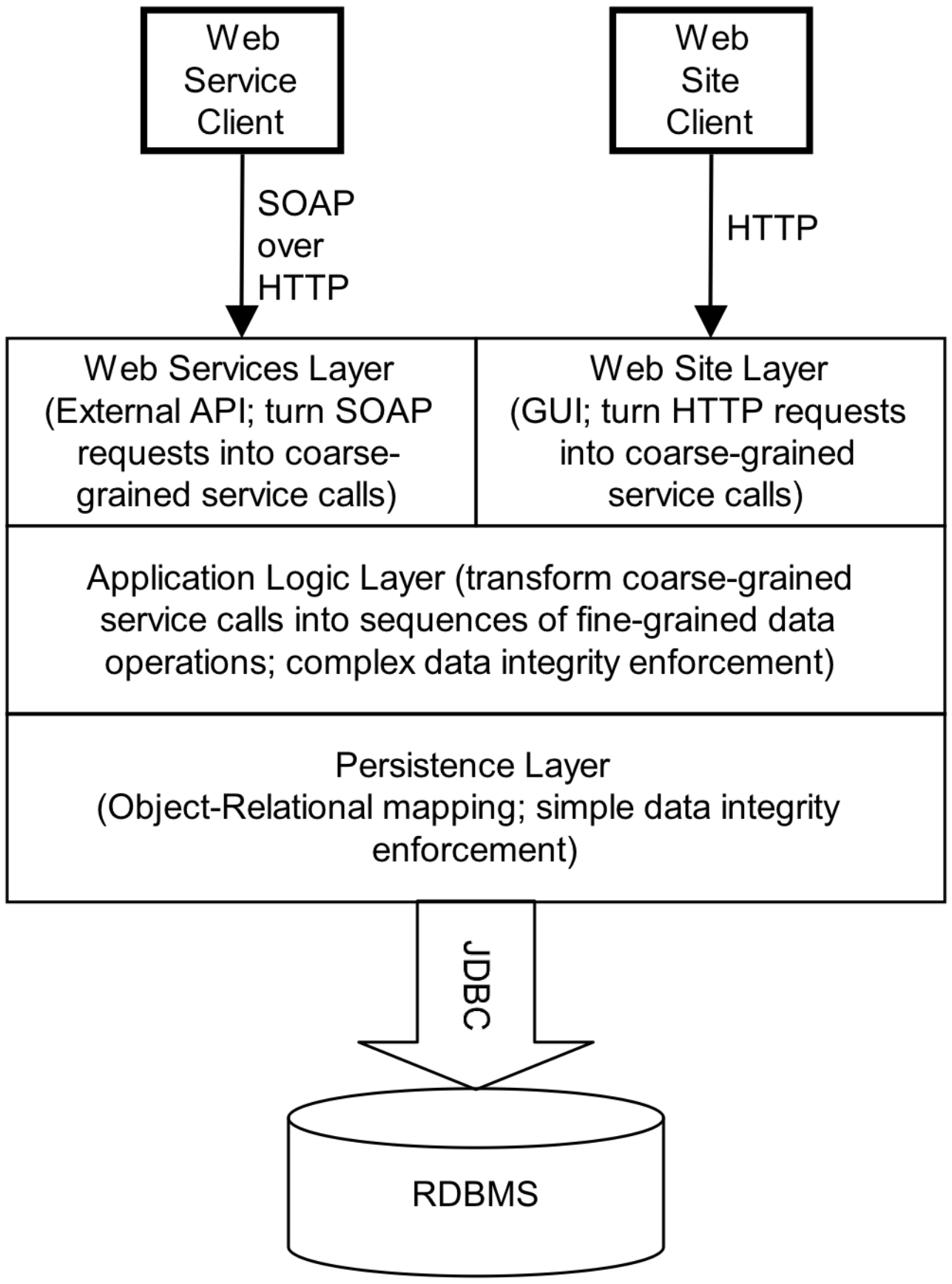

**Figure 4. Data flow layers in the CondorJ2 Architecture**

## 4.2 Example of CondorJ2's Data Centricity

To better understand how the CondorJ2 architecture effectively changes cluster management from a "systems" problem into a "data" problem, this section provides a detailed comparison of the steps required by each system to shepherd a job from submission through successful execution.

[1] Do not interpret this to mean that there must be an instantiated entity bean object residing in memory on the application server for each tuple in the database at any given time - there need not be. The subset of objects actually instantiated at any given time depends on the specific requests the system is processing.

#### 4.2.1 The Condor Approach

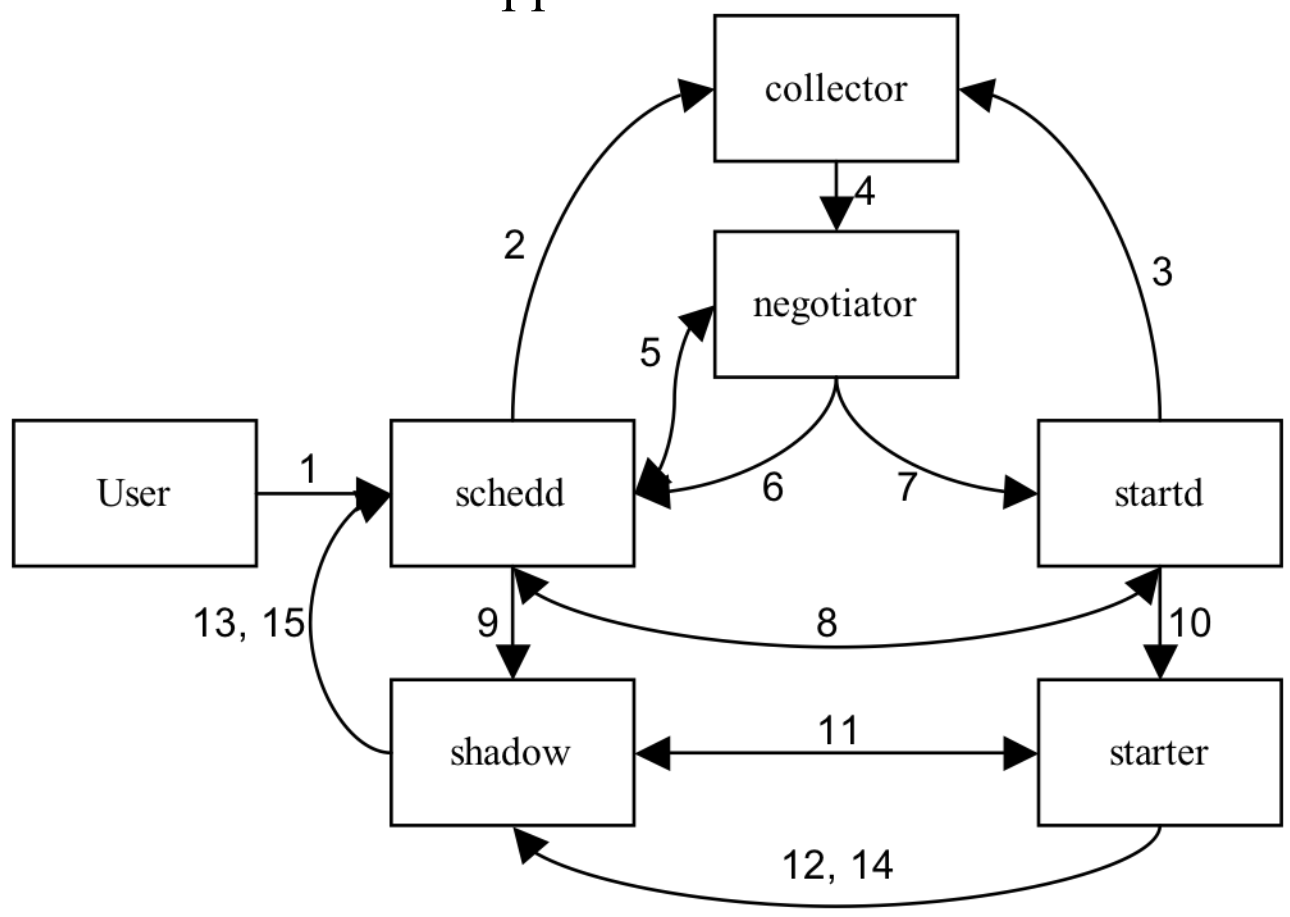

**Figure 5. Data flowing through distributed components in the Condor system**

Figure 5 shows the data flow involved in shepherding a job through the Condor system from submission to completion. Table 1 below describes what happens at each step in the diagram.

**Table 1. Description of Figure 5 steps.**

| Step | Description |
|---|---|
| 1 | User submits job to *schedd*, *schedd* creates job in in-memory queue, logs job to disk |
| 2 | *Schedd* sends job queue summary to *collector* |
| 3 | *Startd* sends periodic heartbeat to *collector* |
| 4 | *Collector* forwards job, machine data to *negotiator* for scheduling algorithm |
| 5 | *Negotiator* contacts *schedd* for job-specific information, *schedd* sends job data to *negotiator* |
| 6 | *Negotiator* informs *schedd* of job-machine match |
| 7 | *Negotiator* informs *startd* of job-machine match |
| 8 | *Schedd*, contacts *startd* to confirm match |
| 9 | *Schedd* spawns *shadow* to monitor job progress |
| 10 | *Startd* spawns *starter* to start up, monitor job |
| 11 | *Shadow, starter* establish socket connection to exchange job state information |
| 12 | *Starter* sends *shadow* periodic job state update messages |
| 13 | *Shadow* forwards job update messages to *schedd* |
| 14 | *Starter* notifies *shadow* when job completes, exits |
| 15 | *Shadow* exits, *schedd* captures exit code, removes job from queue |

#### 4.2.2 The CondorJ2 Approach

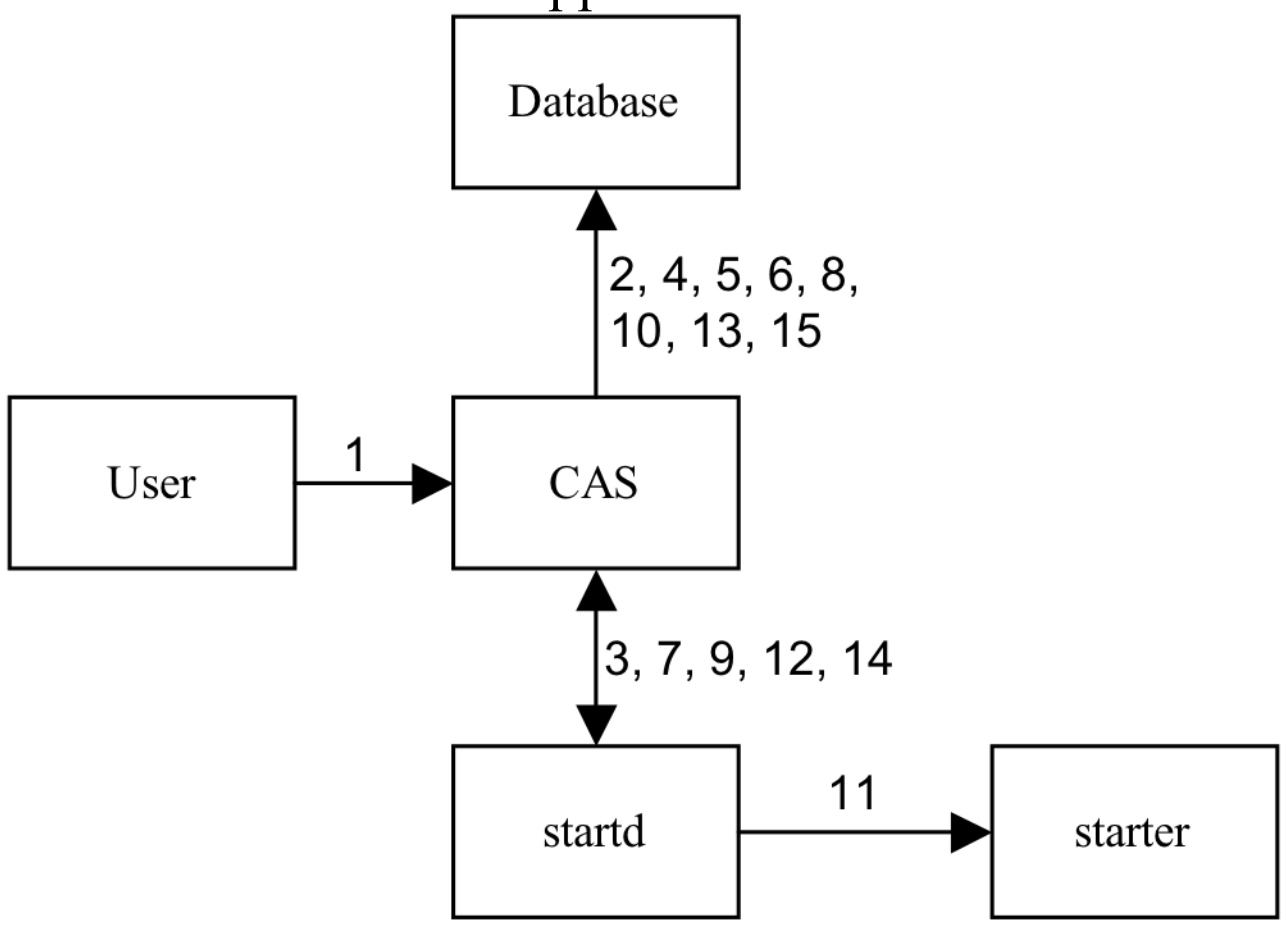

**Figure 6. Data flowing through distributed components in the CondorJ2 system**

Figure 6 shows the data flow involved in shepherding a job through the CondorJ2 system from submission to completion. Table 2 below describes what happens at each step in the diagram.

**Table 2. Description of Figure 6 steps.**

| Step | Description |
|---|---|
| 1 | User invokes submit job service on CAS |
| 2 | CAS inserts a job tuple into database |
| 3 | *Startd* invokes periodic heartbeat web service on CAS |
| 4 | CAS updates a machine tuple in the database, responds OK to *startd* |
| 5 | CAS selects relevant machine tuples, job tuples from database for scheduling algorithm |
| 6 | CAS inserts match tuple, updates related job tuple in db |
| 7 | *Startd* invokes periodic heartbeat web service on CAS |
| 8 | CAS updates machine tuple in database, selects related match and job tuples, responds MATCHINFO to *startd* |
| 9 | *Startd* invokes acceptMatch web service on CAS |
| 10 | CAS deletes match tuple, inserts run tuple, updates related job tuple in the database, responds OK to *startd* |
| 11 | *Startd* spawns *starter* |
| 12 | *Startd* invokes periodic heartbeat web service on CAS, includes job information from *starter* in SOAP message |
| 13 | CAS updates machine tuple, related job tuple in database, responds OK to *startd* |
| 14 | *Startd* invokes periodic heartbeat web service on CAS, includes job completion information in SOAP message |
| 15 | CAS updates machine tuple, deletes related run and job tuples from database, responds OK to *startd* |

#### 4.2.3 Process Management vs. Data Management

While the steps undertaken by each system are conceptually similar, the data flows are quite different. The process-oriented design of Condor results in ten different communication channels between seven distinct entities (six daemon processes and the user) for moving data through the system. Additionally, recall from Section 2 that all operational data is maintained by the daemons in in-memory data structures[2]. This means that accessing or updating a particular piece of data requires communicating with the daemon responsible for managing that data. Considering these factors in combination with the distributed nature of the system, it is not difficult to comprehend why the work involved in building, enhancing and extending these types of systems has typically focused on issues such as OS process management/migration and inter-process communication.

Compare this with the CondorJ2 approach that requires only four communication channels between five entities for moving data through the system. As pictured in Figure 6 and described in Table 2, the focal point of the entire communication flow is the Application Server whose most basic system function is *to transform HTTP requests into SQL statements*. The fact that the core function of the application server is to perform a data transformation is why the work involved in building, enhancing and extending the CondorJ2 system focuses on issues like schema design, concurrency and transaction semantics.

The preceding point is essentially the central claim of this paper, (i.e., that CondorJ2 transforms batch computing from a process management problem into a data management problem) so it is worthwhile to spend an extra paragraph discussing and justifying it. First, since the "live" operational data resides in the database, the system extensibility problem reduces to a data-modeling/schema design problem. The most intellectually demanding task is determining the necessary schema adjustments; it is then possible to "pull" the data up through the three system layers and expose the required service endpoint interfaces. Second, with respect to overall system scalability and performance, the critical factors are the level of concurrency supported by the Application Server and database and the speed and efficiency with which the Application Server can perform the HTTP-to-SQL transformation and the database can process the SQL statements. Typical data management concerns such as a good schema, efficient transformations and short-running transactions for the most common operations are the keys to high performance and improved scalability.

#### *4.2.3.1 Code-base Size*

One additional point to consider when comparing these two divergent approaches is what they translate into in terms of the size of the system's code-base. Since the two systems do not provide a specific, standardized set of services, it is impossible to do an exact "apples-to-apples" comparison. Nevertheless, some rough numbers provided with the appropriate caveats are still interesting to discuss. All numbers refer to source code files only (i.e., no make files or build scripts) and include comment lines.

The total Condor code-base is about 470,000 lines. Much of this code consists of support classes and libraries that are used by multiple system components. It would be very difficult (if not impossible) to determine how much of this code is used to provide services that both systems provide. It is, however, possible to estimate the amount of code that is clearly specific to system components that do provide services common to both systems. We estimate that about 69,000 lines of code fit this description[3].

The total CondorJ2 code-base is about 62,000 lines. Of this we estimate that about 35,500 lines of code are specific to common services. The remainder of the CondorJ2 code-base consists of configuration management (~11,000 lines; operational and historical), historical machine information management (~9,000 lines) and the web GUI and support classes (~6,500 lines).

## 5. EXPERIMENTAL PERFORMANCE ANALYSIS

The purpose of this section is to gain an understanding of how the prototype CondorJ2 implementation performs under a variety of operating conditions. More specifically, we present here the experimental results that give insight into the system's ability to:

- Achieve high scheduling throughput
- Manage a large number of machines
- Schedule a workload of mixed-length jobs

Section 5.1 provides a description of these metrics as well as a justification for why we believe they are important. Section 5.2 presents results from several experiments we performed on the CondorJ2 prototype. Finally, to place the CondorJ2 findings in context, Section 5.3 presents the results from measuring Condor's performance for the same three metrics. During the course of experimentation we discovered many things that merit a more thorough treatment than that provided here. Since the goal of this paper is to introduce the CondorJ2 system, though, we limit our discussion to a high-level overview of Condor's performance.

Before proceeding, an important point to note is the distinction in Condor (and CondorJ2) between *physical* machines and *virtual* machines. In both systems, a single physical machine is home to one or more virtual machines. In both systems, scheduling decisions are made at the virtual machine, not the physical machine, level. Thus, a given physical machine can be simultaneously executing one job for each virtual machine resident on it. Note that the use of the term *virtual* serves only to signify a level of abstraction in the way the system models execute nodes – it does not imply multiple separate operating systems and process spaces on the actual nodes[4].

The number of virtual machines on any given node is a configurable parameter with the default being one per processor. For our experiments we only had 50 physical machines available for use and this was not enough for many of the experiments we wished to perform. To get around this limitation we changed the virtual-machine-to-physical-machine ratios in order to simulate a variety of cluster sizes. For most of our experiments the fact that we have more virtual machines than actual processors makes no difference. For those cases in which this simulation was not transparent, the effects are discussed as part of the analysis.

[2] The persistent job queue log maintained by the *schedd* is used only for recovery. A user could parse it and replay it to reconstruct the current state of the job queue but this is neither a common nor convenient approach to querying the system.

[3] This excludes the *startd* and *starter*. CondorJ2 currently uses a modified version of the Condor *startd* and *starter*, so those components are about the same size.

[4] All occurrences of the term virtual machine in this paper refer only to the definition provided here. The namespace collision between the Condor notion of a virtual machine and the more common notion of a virtual machine is unfortunate.

## 5.1 Scalability and Performance Metrics

The purpose of this section is to clarify and motivate the metrics we use to characterize performance and scalability in this analysis.

### 5.1.1 Scheduling Throughput

One metric for characterizing the scalability and performance of a cluster management system is the number of jobs that it can schedule per second. In a fully and continuously utilized system with a fixed workload, the **average scheduling throughput demand** is defined by the ratio of the number of execute machines to the average length of a job. A system with 1,200 execute nodes subject to a workload consisting solely of 20-minute jobs, for example, must be capable of a scheduling throughput rate of at least one job per second in order to keep the entire cluster operating at maximum capacity.

Note here that in a fully utilized cluster, whenever a machine finishes executing a job there will always be another eligible job waiting in the queue to take its place (if there is no job to take its place then the cluster is not being fully utilized). One important implication of this is that overall system throughput is affected not only by the time it takes to make scheduling decisions and start up jobs, but also by the efficiency with which the system can perform any necessary post-execution processing. Post-execution tasks include recording historical information about the job, recording accounting information and removing the job from the queue.

Because the average scheduling throughput demand is a ratio, it can be driven upwards by either increasing the numerator (number of execute machines) or decreasing the denominator (average job length). In practice, two well-established trends - i.e., machines keep getting both faster and cheaper - suggest that it is not unreasonable to assume that the numerator is increasing at the same time that the denominator is decreasing, further emphasizing the importance of scheduling throughput as a system metric.

### 5.1.2 Managing a Large Cluster

The previous section highlighted the importance of considering scheduling throughput when assessing scalability. It is important to recognize, though, that scheduling throughput is only one factor influencing system scalability. Consider, for example, that a 60-processor cluster running one-minute jobs and a 36,000-processor cluster running 10-hour jobs are both one-job-per-second systems. Clearly, not all one-job-per-second systems are the same.

In the example just given, fixed system overheads are likely to be much higher in the 36,000-processor system than in the 60-processor system. The fact that one cluster is 600 times larger than the other suggests that there will be commensurate increases in the amount of machine information to manage, in the size of the job queue to manage, in the amount of message traffic to process[5], and, presumably, in the number of system-status and job-monitoring queries to respond to. Depending on the system architecture, there may also be increased RAM requirements and an increase in the number of concurrently running job-monitoring processes (e.g., the *shadow*). Thus for large clusters, scalability in terms of the absolute number of machines is still an important concern even if the workload generates a low job turnover rate.

[5] Even if the scheduling rate is the same, the nodes still need to communicate with the scheduler and job queue manager periodically during the course of the job to make sure the job is not dropped, has not been removed by the user, etc.

### 5.1.3 Scheduling Mixed Workloads

To this point, the scheduling throughput rate for a given system has been treated as a static function of cluster size and average job execution time. This works provided that the mix of jobs in the system is always the same and that the system has the ability to schedule jobs in such a way that the turnover rate is maintained at the constant average value. In general, this is not always the case.

Consider, for example, a workflow consisting of 960 one-minute jobs and 240 six-minute jobs. The total execution time required to complete the workflow is 2,400 minutes with an average job execution time of two minutes. For a 120-machine cluster, the two-minute average implies an overall scheduling throughput requirement of one job per second sustained for twenty minutes. In order for a one-job-per-second scheduler to maintain this rate, it must be able to generate an intelligent schedule that interleaves the jobs in such a way that the turnover rate remains constant[6].

Even if the scheduling algorithm is smart enough to balance turnover rates, the system may not have the freedom to do so. Consider the same workflow described above, but now add in the constraint that the output data generated by the one-minute jobs serves as the input data for the six-minute jobs – i.e., the six minute jobs cannot begin until the one-minute jobs have completed. Now the twenty minutes of one job per second throughput breaks down into eight-minutes of two jobs per second followed by 12-minutes of 1/3 job per second. The *average* throughput on the workflow is still just one job per second, but the underlying job mix requires a capacity of two jobs per second for the cluster to operate at maximum capacity. The cluster management system must be able to handle the skew induced by a dynamic workload to consistently reach peak efficiency.

## 5.2 CondorJ2 Experiments

This section begins with an overview of the general experimental setup followed by a presentation and discussion of results.

The daemons on the execute nodes are the Condor version 6.7.x *startd* and *starter* modified to communicate with the CAS using the gSOAP [12] library. We used IBM's DB2 Universal Database version 8.2 and JBoss AS version 4.0.4 in all experiments. Both the application server and DBMS are running on a single physical machine - a 3.0 GHz Quad-Xeon server with 4.0 GB of RAM and 1TB of disk space in a RAID-5 configuration. According to *hdparm*, the disk is capable of cached reads of approximately 1740 MB/sec and buffered reads of approximately 125 MB/sec.

### 5.2.1 Scheduling Throughput

Execute nodes in CondorJ2 always the initiate any interaction they have with the CAS. One consequence of this is that CondorJ2 follows a pull model for job scheduling in the sense that the execute nodes "pull" jobs from the server-resident queue(s). This means that there is no knob to turn or parameter to set that controls the rate at which the server hands out jobs to the execute nodes. Rather, since the server is always reacting to client-generated events, we must increase the rate at which the execute nodes request jobs in order to see how the CAS responds to an increased demand for scheduling throughput.

[6] Neither Condor nor CondorJ2 has this capability.

To accomplish this we simulated a 180-node cluster by configuring 45 physical machines to act as four virtual machines each. We then pre-loaded the system with a number of identical, fixed-length jobs sufficient to maintain the desired throughput rate for at least twenty minutes. We conducted five separate experiments in which the length of the pre-loaded jobs was varied in order to measure system performance for five different rates. The job lengths ranged from a minimum of six seconds to a maximum of five minutes in order to cover a range from 30 jobs per second down to 0.6 jobs per second.

Figure 7 plots the number of jobs scheduled per second as a function of the length of the jobs. The top line shows the ideal throughput rate required to keep the entire cluster running at maximum capacity while the bottom line shows the results we observed in our experiments. For the five-minute, one-minute and eighteen-second jobs, we observed the CAS achieving scheduling throughput rates very close to the maximum. For the nine-second and six-second jobs the observed rate is below the maximum.

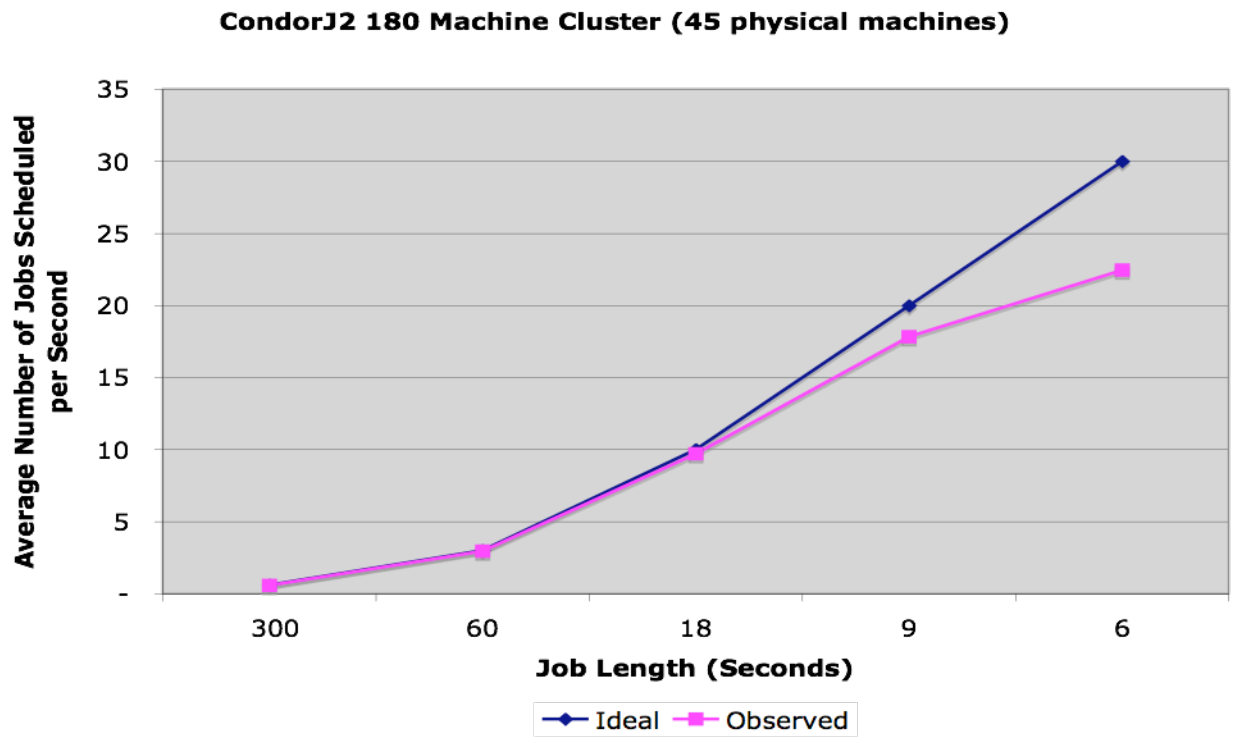

**Figure 7. Scheduling Throughput vs. Job Length in CondorJ2**

One possible explanation for this rate reduction is that the CAS becomes overwhelmed somewhere in the range between ten and twenty jobs per second and is not able to meet the generated demand. This seems unlikely, though, when we consider that in the six-second-job experiment we observed the CAS maintaining a throughput rate greater than twenty jobs per second.

Given that (due to resource limitations) many of the machines in our test cluster are rather slow (they are a mix of single and dual processor 1GHz P3 machines), an alternative explanation is that some of the execute nodes are overwhelmed by the task of churning through so many jobs so quickly and, consequently, are preventing the cluster as a whole from generating the desired throughput rate. Before we can decide to accept this explanation we require two pieces of additional evidence. First, we need evidence that the execute nodes are encountering problems that degrade performance. Second, we need evidence that the CAS has excess capacity at the highest observed throughput levels.

One common problem that occurs in distributed computing is that an execute node encounters some error condition that causes it to fail to run (i.e., “drop”) a job[7]. Figure 8 shows a bar chart plotting the number of distinct execute nodes that dropped at least one job during runs of the experiment. The leftmost series is the five-minute jobs and the rightmost series is the six-second jobs. The left-hand bar in each series counts the number of distinct *virtual* nodes that dropped at least one job. The right-hand bar counts the number of distinct *physical* nodes that dropped a job. As is clear from the figure, very few nodes encountered problems when running the one and five minute jobs. With the eighteen-second jobs, some of the nodes encountered problems, though not enough to materially affect the observed throughput rate. For the shortest jobs, though, significant portions of the cluster are encountering errors. For the six-second jobs, almost 40% of the virtual nodes encountered problems, and *every single physical node* was home to at least one virtual node that that dropped a job.

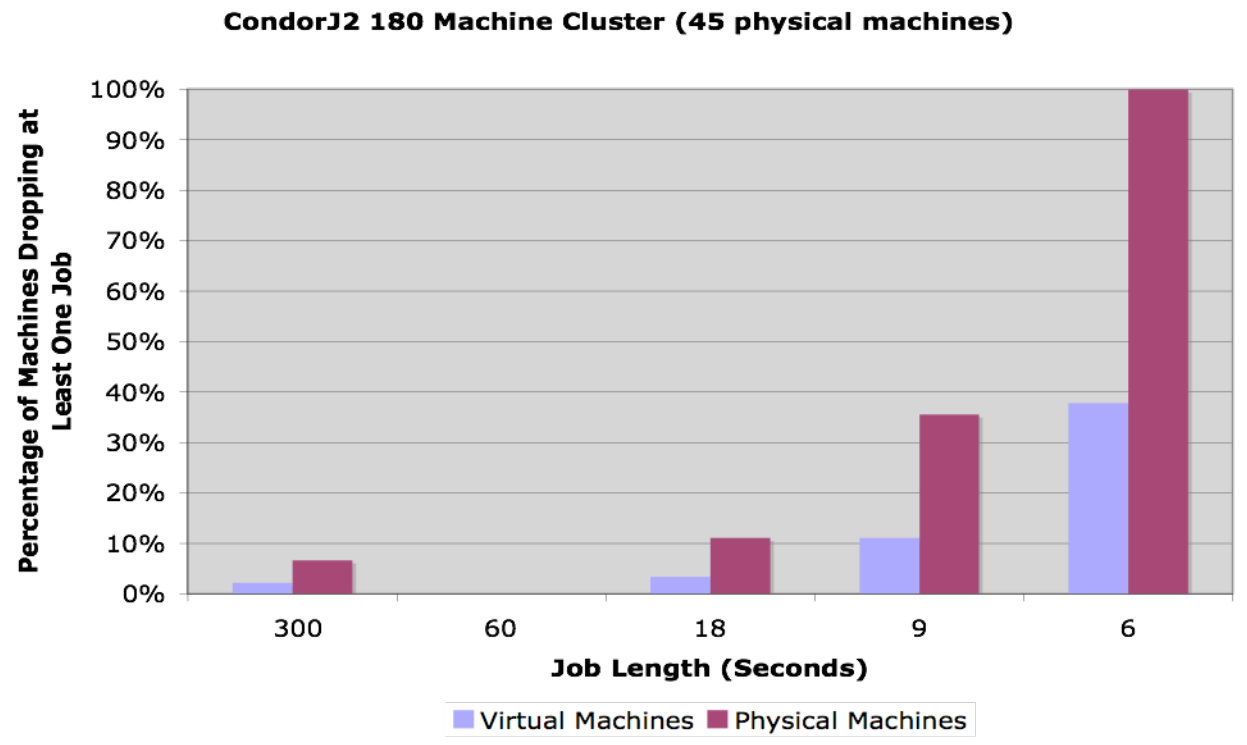

**Figure 8. Execute Hosts Failing to Run Jobs**

Figure 9 shows a plot of CAS CPU cycle consumption as a function of the number of jobs scheduled per second. The CPU usage information was collected during the experiments by a process on the CAS that woke up once every minute and pulled statistics from the /proc file system (these are the same statistics that *top* uses). To produce the points, for each experiment we calculated the average scheduling throughput rate, excluding the ramp up and ramp down time. Since the CAS records the timestamp of all the relevant events, we were able to correlate these rates with the appropriate CPU usage statistics.

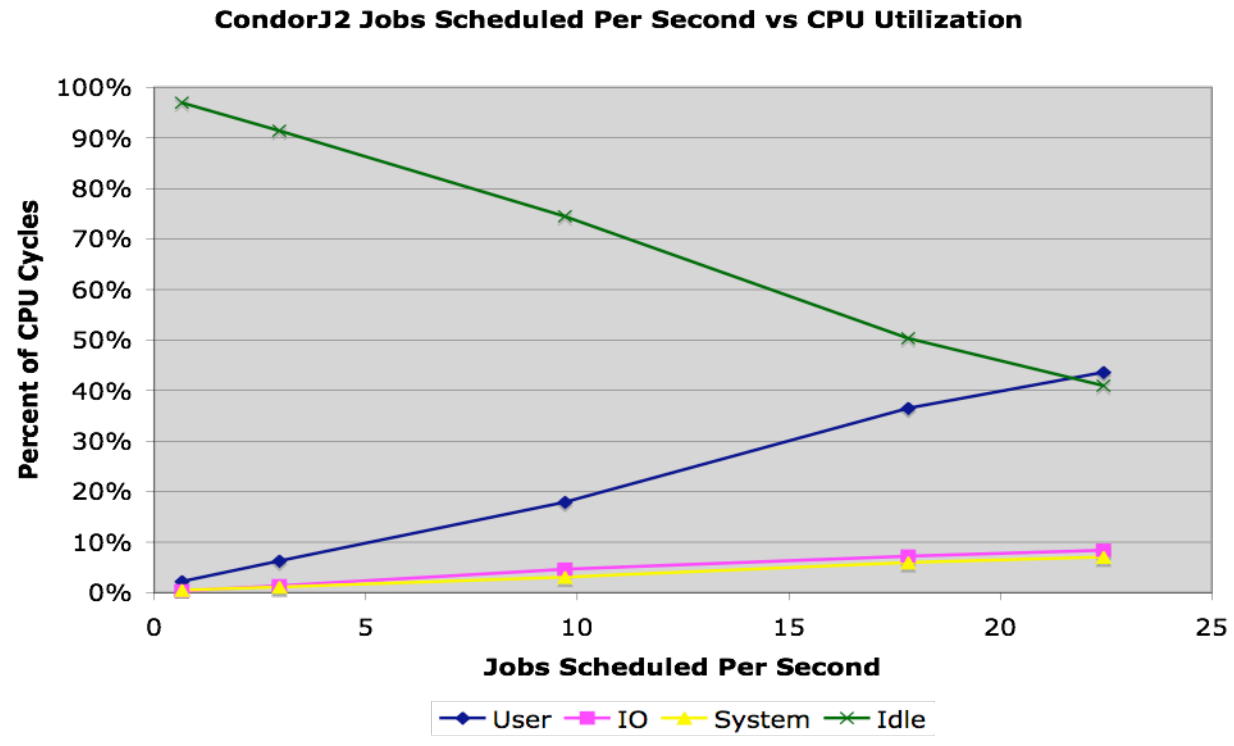

**Figure 9. CAS CPU Utilization vs. Scheduling Throughput**

The bottom two lines in Figure 9 are IO usage (cycles spent waiting for the disk) and System usage (cycles spent executing in “kernel” mode). The middle line is User usage (cycles spent doing actual computation) and the top line is Idle cycles (spare computational capacity). All four categories sum up to 100%. As Figure 9 shows, the CAS still had significant capacity to spare during all experiments, lending additional credence to the hypothesis that observed scheduling throughput is being limited by errors on the execute nodes and not by the server.

[7] Ensuring that the *job queue manager* does not drop jobs is one reason why job management requires transactions.

One striking feature of Figure 9 is the apparent linear growth in cycle usage in response to increases in scheduling throughput. Also noteworthy is the fact that cycles spent on computation grow at a much faster rate than those spent on IO or System tasks. We believe that JBoss is consuming more of the User cycles than DB2, though we did not collect the process-level information necessary to substantiate that claim.

To wrap up this section, it seems appropriate to address why the nodes are dropping jobs. A quick glance through the log files turned up numerous "timeout" errors. This suggests that setting up and tearing down the environment for running jobs at the rate of four jobs (one per virtual machine) every six seconds is not sustainable for our test-bed nodes. Fortunately, this is unlikely to be a problem in a "real" cluster environment; still, it is disappointing that resource limitations prevented us from obtaining cleaner results.

### 5.2.2 Managing a Large Cluster

As discussed in Section 5.1.2, with a workload of sufficiently long running jobs, even very large clusters can have relatively modest scheduling throughput requirements. In these situations it is possible for cluster size dependent overheads to become a barrier to scalability. To better understand the level of size dependent overhead imposed by CondorJ2 we conducted an experiment in which we simulated a 10,000-node cluster by configuring 50 physical machines to manage 200 virtual machines each. We ramped up to full utilization by turning on the execute nodes and then submitting the jobs into the system in 20 batches of 2,500 jobs each at five minute intervals. To avoid overwhelming the execute nodes by trying to start up too many jobs too quickly, each batch targeted five percent of the virtual machines (10 per physical machine) so total ramp up time was approximately 100 minutes. This extended ramp up time also kept the job turnover rate low enabling a better focus on the size-dependent overheads.

Figure 10 plots eight hours of CAS CPU utilization statistics versus elapsed time during this experiment. The utilization statistics plotted here are five-minute rolling averages. As in Figure 9, the IO and System cycles are on the bottom, User cycles are in the middle and Idle cycles are on top.

Figure 10 has several noteworthy features. The first is the spike in User and System cycle usage that occurs at the beginning of the experiment. Part of this is due to the one-time startup costs for things like creating database connections, filling caches, allocating bean instances, etc. Another factor contributing to this spike is the fact that whenever an execute machine restarts, the CAS monitors and records extra historical information about machine attributes that only change when the machine is rebooted (e.g., operating system, architecture, total physical memory, etc.). When 10,000 virtual machines all restart in a short period of time this extra workload can quickly add up.

Another interesting feature of Figure 10 is the apparent oscillation between sustained periods higher and lower cycle consumption. The high plateaus represent periods when jobs are completing and new jobs are being scheduled for execution. Because there are 10,000 machines in the cluster, and because the jobs were submitted in twenty batches at five minute intervals, the high plateaus last approximately 100 minutes each and correspond to a scheduling throughput rate of ~1.67 jobs/sec (10,000 jobs scheduled every 6000 seconds). The low plateaus represent periods in which no jobs are turning over so server-side workload consists almost entirely of handling periodic heartbeat messages from the execute nodes. The jobs are 150 minutes long so these plateaus between periods of job turnover last about 50 minutes.

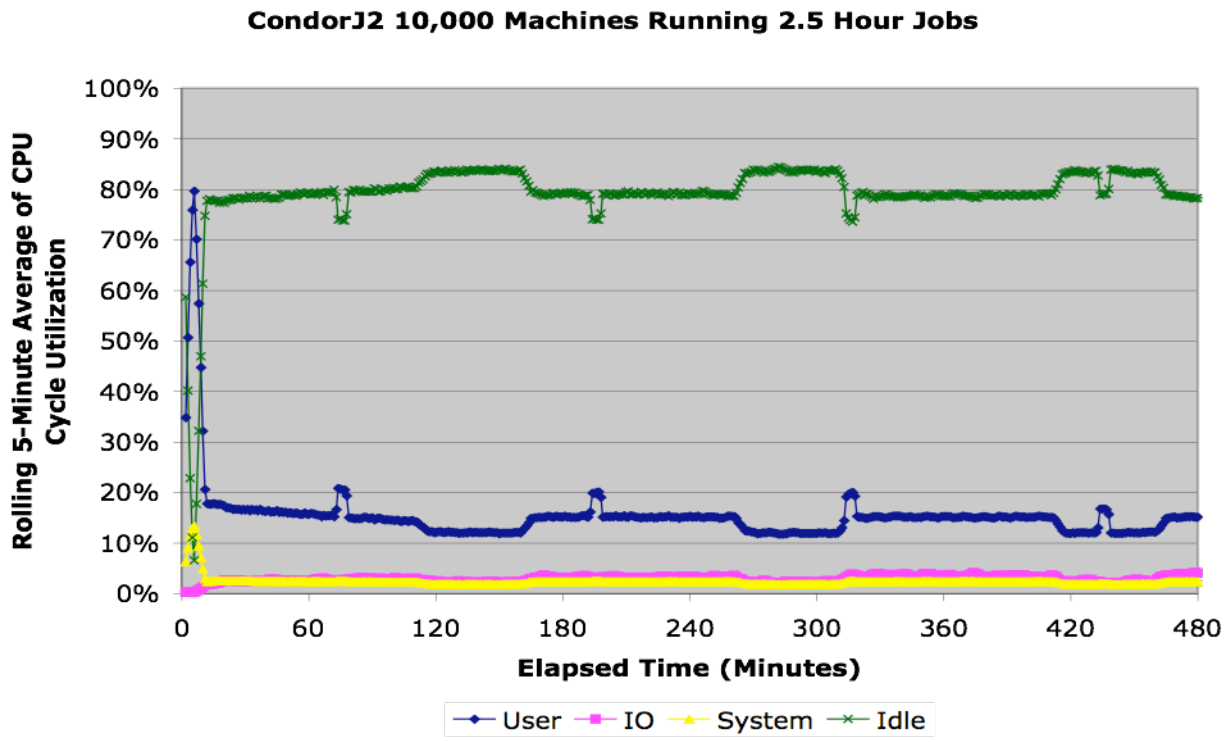

**Figure 10. CAS CPU Utilization in a 10,000 Virtual Machine Cluster**

Finally, Figure 10 shows four distinct spikes at almost exactly two-hour intervals. These spikes result from a DB2 background process. Because these spikes do not significantly affect system performance, we did not investigate whether they represent checkpointing, statistics collection or some other periodic action.

The fact that the CAS has a significant amount of spare capacity in this experiment suggests that it could support more than 10,000 machines at such a low scheduling throughput rate. Unfortunately we could not test this hypothesis because our test-bed machines began to encounter memory allocation problems when we tried to configure them to have more than 200 virtual machines each. Since we only had 50 machines available for experimentation, this capped the largest cluster we could simulate at 10,000 machines.

### 5.2.3 Scheduling Mixed Workloads

Section 5.1.3 discussed how skew in the workload can lead to suboptimal cluster utilization even if the average job turnover rate is below the cluster management system's overall scheduling throughput capability. This section presents the results of experiments that we ran in order to understand CondorJ2's ability to cope with a mixed workload. Though we ran several sets of these experiments with varying cluster sizes and job queue lengths, the presentation here focuses on only one experiment (the largest scale one that we ran) because there were no substantial differences across the various combinations that we tried.

In this experiment we configured 45 physical machines to manage twelve virtual machines each in order to emulate a 540-machine cluster. We loaded the job queue with 6,480 one-minute jobs and 1,620 six-minute jobs. Total workload is therefore 16,200 minutes for 8,100 jobs resulting in an *average* job runtime of two minutes per job. With a 540-machine cluster this implies an optimal time to completion of 30 minutes and an *average* scheduling throughput demand of 4.5 jobs/sec. There are no inter-job dependencies, so the system can schedule jobs in any order.

Figure 11 plots elapsed time (minutes, horizontal axis) against the number of jobs in progress (vertical axis) during this experiment. The graph shows that by the end of the second minute, the system is running at full capacity with all 540 nodes executing a job. The system stays at full capacity until, the $32^{nd}$ minute when all of the jobs have completed. The slight dips in the graph (e.g., at minutes 3, 12 and 25) occur when the lag between the time that a virtual machine reports completing one job and beginning the next job happens to span a minute boundary.

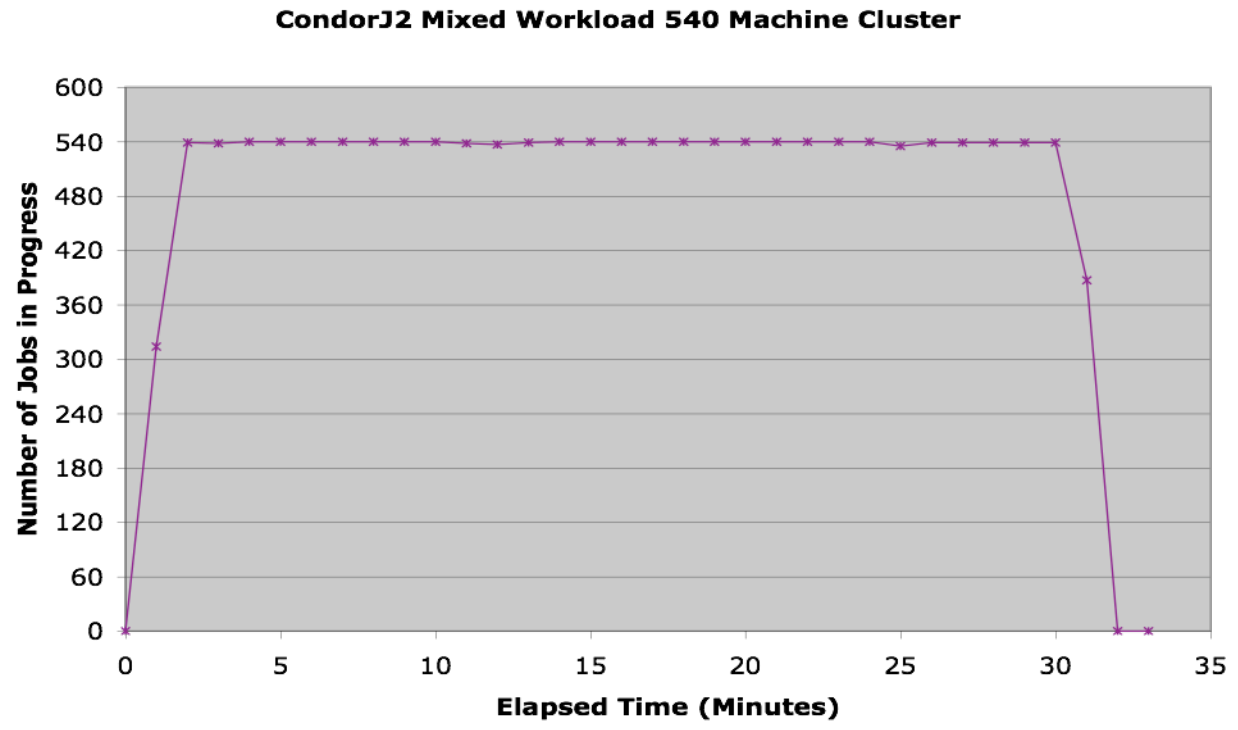

**Figure 11. CondorJ2 Mixed Workload Scheduling**

Figure 12 is useful for understanding how CondorJ2 copes with the mixed workload. The horizontal axis measures elapsed time while the vertical axis measures the job completion rate. There is a two-minute ramp up period followed by an approximately twelve-minute period in which jobs are turning over at a rate of almost nine jobs per second (dropping to six during the 14th minute). During this period CondorJ2 is working its way through all of the one-minute jobs. The nine jobs per second rate follows from the fact that each of the 540 nodes is turning over one job per minute. The twelve-minute duration follows from the fact that there are 6,480 of these jobs and 540 nodes to run them on.

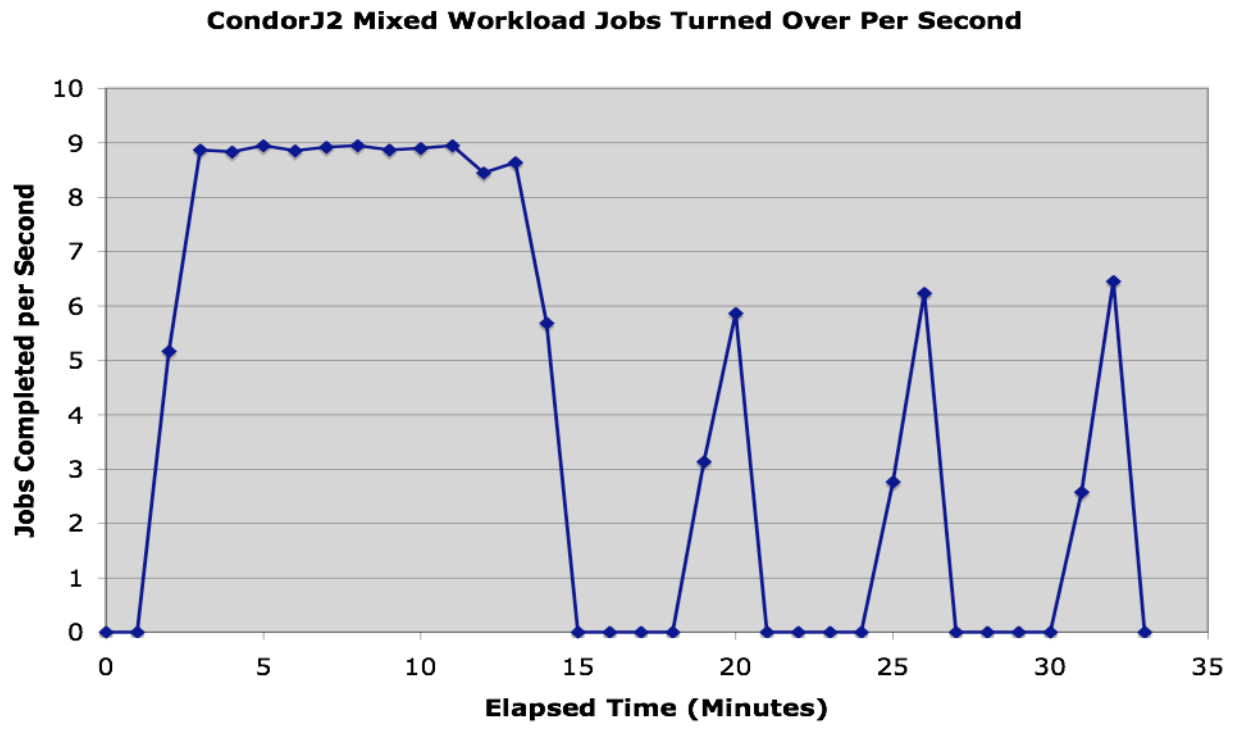

**Figure 12. CondorJ2 Mixed Workload Job Turnover Rate**

The next segment of the graph contains an alternating pattern of lulls in which no jobs turnover and spikes in which jobs turnover at a rate of at first three and then six jobs per second. During this period CondorJ2 is working through all of the six-minute jobs. The six-minute lag between peaks occurs because the jobs are six minutes long. The rate fluctuation between three and six jobs per second is deceiving. There is actually a 60 second span in which jobs are turning over at a rate of nine per second but, because it crosses minute boundaries, it appears as two separate rates that sum to nine. Recording job completion and generating the new job-machine match is obviously not instantaneous. This small lag seems to be the most likely reason that the three peaks creep slightly higher during the course of the experiment.

As Figure 12 illustrates, CondorJ2 copes with the mixed workload with a "brute-force" approach. There is no specialized scheduling algorithm here to smooth out the turnover rate. Such an algorithm is unnecessary in this example because the throughput rate is low enough that the CAS can handle the workload without one.

## 5.3 Condor Experiments

The previous section analyzed CondorJ2 system performance on three different metrics. The purpose of this section is to place those findings in context by performing a similar analysis on Condor. The focus of this paper is on CondorJ2, so the Condor analysis presented here is less detailed than the previous analysis.

In all of the Condor experiments we used the same hardware as we used in the CondorJ2 experiments. The execute nodes are drawn from the same pool of fifty machines while the "server-side" components (e.g., the *collector*, *negotiator*, *schedd* and *shadow*) run on the same Quad Xeon that CondorJ2 used for JBoss and DB2. Since the *schedd* is single-threaded, in some experiments we ran three *schedds*[8] on the machine simultaneously (and split the jobs up accordingly) to give Condor the opportunity to take advantage of the extra CPUs. Experiments in which we took this approach are noted as necessary. We used Condor version 6.8.2, the latest stable release, in all experiments.

### 5.3.1 Scheduling Throughput

Recall from Section 2.2 that Condor uses a push model for moving jobs from the queue to the execute hosts. The Condor push model has two phases. In the first phase, the *negotiator* identifies matches between jobs and machines and notifies the relevant *starter* and *schedd*. In the second phase the *schedd* contacts the *starter* and passes it the job information. Under certain circumstances[9] the *negotiator* can be bypassed entirely, resulting in higher scheduling throughput. In these situations the *schedd* alone assumes responsibility for turning over the jobs in its queue. Since this is both a common case and the best performance case for Condor, this is the case we focused our experiments on. Consequently, this analysis ignores *negotiator* costs and looks exclusively at *schedd* performance.

Reflecting the fact that it pushes jobs out to the execute nodes, the *schedd* has a configurable parameter referred to as the "job throttle". The job throttle parameter sets an upper bound on the number of jobs per second that the *schedd* will attempt to start up. The default value for the job throttle is one job every two seconds. The Condor manual cautions against increasing the throttle past the default value lest the *schedd* become overwhelmed. Simply citing this figure is a rather unsatisfactory way of measuring Condor, so we chose to run some experiments to see if we could determine an upper bound on the actual throughput rate that our hardware could support for a single *schedd*. (Section 5.3.3 contains some insights into how scheduling throughput is affected by distributing the job queue across multiple *schedds*.)

We conducted a series of experiments to try to measure the rate at which the *schedd* could turn over jobs. We varied the job throttle setting across experiments so that we could observe how the system responded to a range of throughput demands. In each experiment we preloaded the job queue with a collection of one-minute jobs and simulated a cluster with enough virtual machines to keep the *schedd* busy at the desired rate; to test at five jobs per second, for example, we set up a cluster with 300 virtual machines. Initially we were confused by the results because the observed scheduling rate varied throughout the course of any given experiment. Eventually we found that there is a relationship between observed scheduling throughput and job queue length.

[8] We reserved the fourth cpu for the other system processes.

[9] E.g., when the *starter* completes a job and the *schedd* has a substantially similar job sitting idle in the queue.

Figure 13 shows a plot of the observed scheduling throughput against the number of jobs in the queue when the job throttle is set at two jobs per second. As indicated by the graph, scheduling throughput begins to drop below two jobs per second once there are approximately 1,800 jobs in the queue. When there are at least 5,000 jobs in the queue, scheduling throughput drops below one job per second. (The two points on the extreme left can be ignored because they represent times when there are so few jobs in the queue that there is not enough work to do to maintain the desired throughput rate.) We repeated this experiment at several throttle settings and observed the same basic relationship.

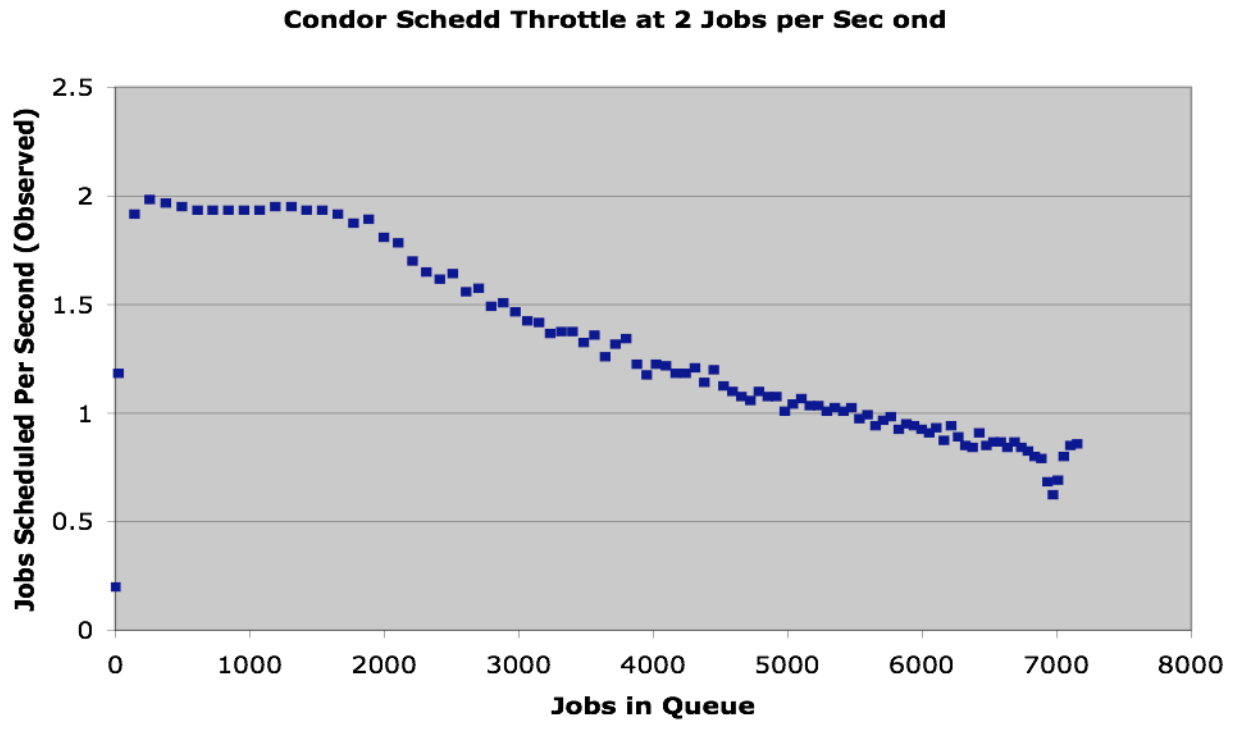

**Figure 13. Condor Scheduling Rate vs. Job Queue Length**

To understand what is happening here, it is instructive to look at Figure 14. Figure 14 contains a plot of *schedd* CPU cycle utilization versus job queue length for the same experiment. Looking at the right hand side of the graph, the top line is User cycles, the middle line is IO cycles and the bottom line is Idle cycles. Because the *schedd* is single-threaded, and because the machine it is running on has four processors, a lone *schedd* can never consume more than 25% of the total available cycles. For clarity, the numbers here have been adjusted. The User and IO numbers have been multiplied by four to better reflect the intuitive notion of when the *schedd* has used all available cycles; the Idle cycles are calculated as 100% less the sum of User and IO usage.

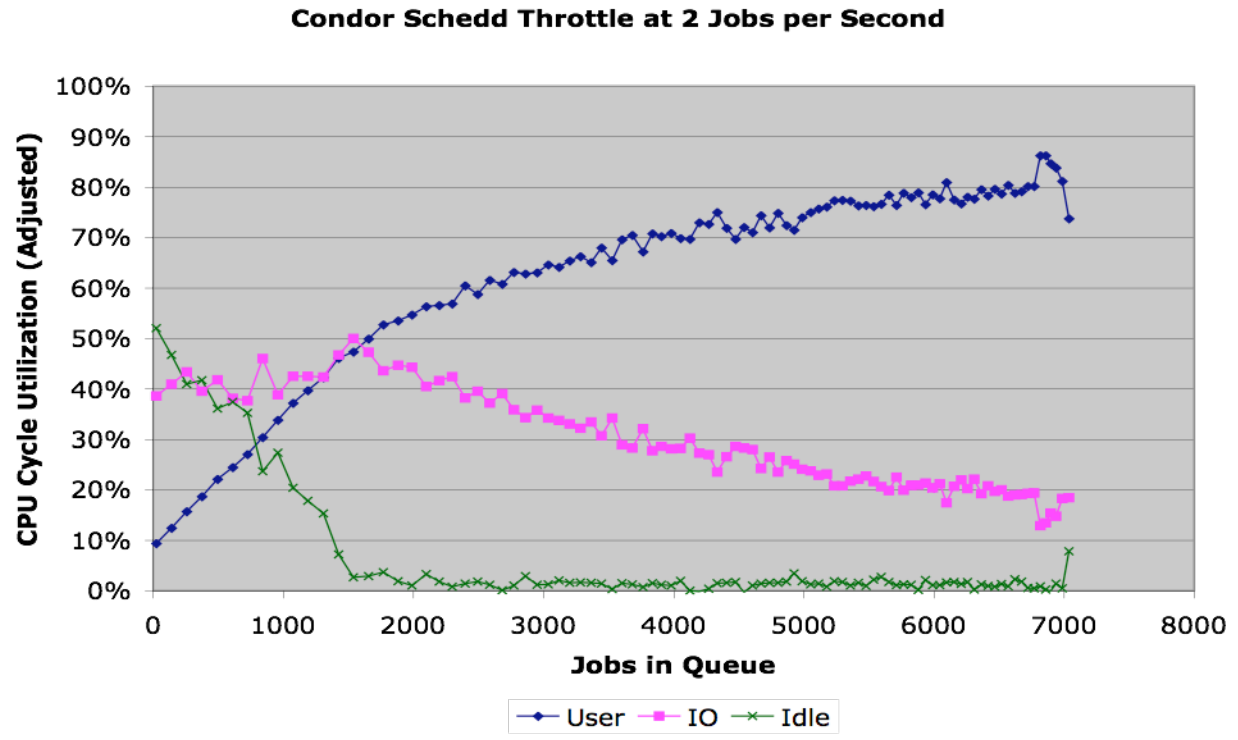

**Figure 14. Condor CPU Usage vs. Job Queue Length**

According to Figure 14, CPU usage increases linearly from 0 to about 2,000 jobs in the queue. After this point, the *schedd* runs out of available cycles leading to damped growth in User cycles and a decrease in IO wait cycles. The point at which the *schedd* maxes out the CPU is close to the point at which job scheduling throughput begins to fall below the throttle rate. Given these factors in combination with the knowledge that the *schedd* has other responsibilities in addition to turning over jobs (e.g., talking to the *negotiator*, responding to queries, etc.), we decided to set the *schedd* throttle at one job per second (double the recommended rate) in our remaining experiments.

### 5.3.2 Managing a Large Cluster

We worked through a number of different approaches to try to get a single *schedd* to manage 5,000 simultaneously running jobs. As with CondorJ2, we pulsed jobs into the system to keep the job turnover rate low and to better focus on the size-dependent overheads. In some attempts we could ramp up to 5,000 jobs in progress, but Condor would crash once the jobs started to turn over. We did not conduct any smaller scale experiments for a single *schedd*. We also did not investigate if multiple *schedd*s running on our hardware could manage a 5,000-node cluster.

### 5.3.3 Scheduling Mixed Workloads

To assess how Condor manages a mixed workload we ran the same set of experiments described in Section 5.2.3.

In the experiments presented here we configured 45 physical machines to run four virtual machines each in order to emulate a 180-node cluster. The workload was 2,160 one-minute jobs and 540 six-minute jobs. For our 180-node cluster this yields an optimal total runtime of 30 minutes at an average throughput of 1.5 jobs per second. It seemed pointless to run the experiments with a single *schedd* with the job throttle set at one job per second just to conclude that a one-job-per-second scheduler cannot effectively manage a mixed workload demanding an average throughput level of 1.5 jobs per second. Instead, we decided to distribute the job queue evenly across three *schedd*s. This gave the *schedd*s a better opportunity to cope with the workload and also gave us the opportunity to see if distributing the job queue across multiple *schedd*s impacts system performance. Since all three *schedd*s have the throttle at one job per second, the system's aggregate throughput capacity exceeds the workload's demand.

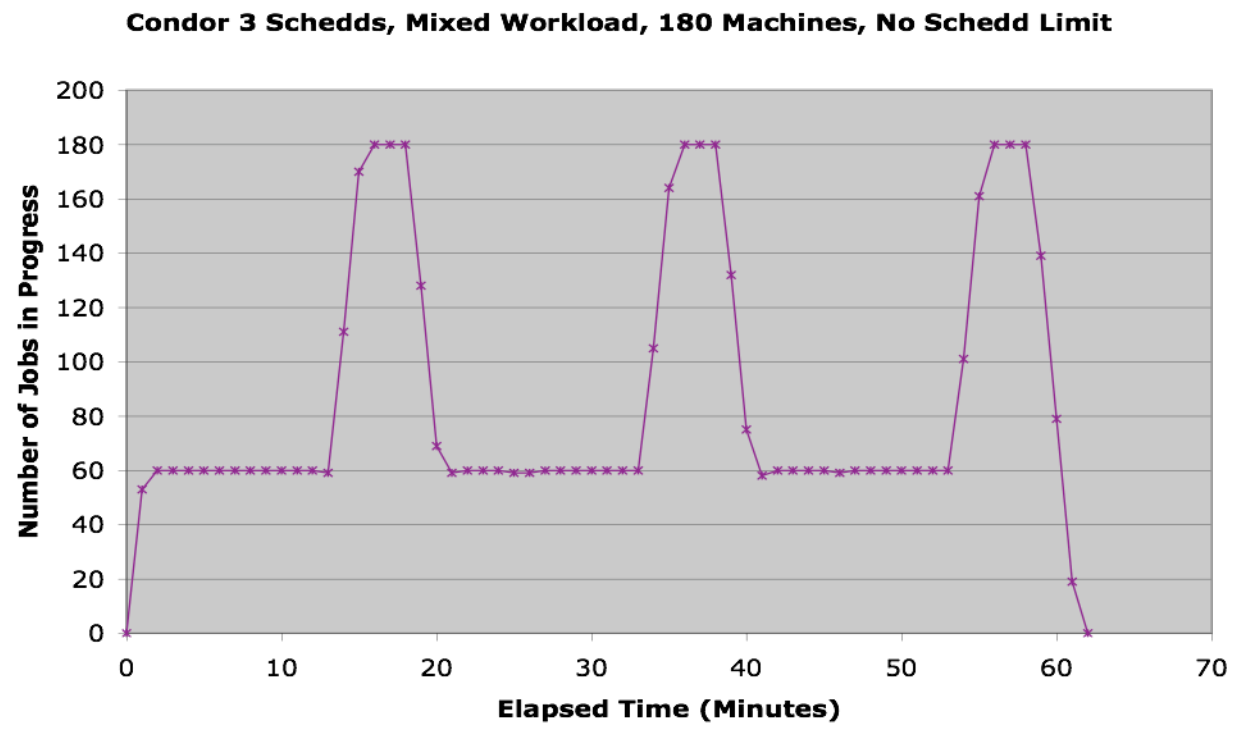

**Figure 15. Condor Mixed Workload, No Schedd Limit**

Figure 15 shows the number of jobs in progress plotted against elapsed time for one run of this experiment. What is happening here is that the *negotiator* begins by picking one *schedd* and allocating all 180 machines to it until it drains its queue. It repeats this for the second and third *schedd*s at which point the entire workload is completed. Because each *schedd* is limited by the job throttle to starting up one job per second, they are only able to simultaneously manage 60 one-minute jobs; the *schedd* maintains a claim on the other 120 nodes in the cluster, but those nodes sit idle until the *schedd* has completed all of the one-minute jobs and starts scheduling the six-minute jobs. At this point it can ramp all the way up to 180 machines. The result of this behavior is that the cluster is underutilized and the 30-minute workload actually takes about 60 minutes to complete. One way to circumvent this problem in Condor is to configure the *schedd*s to maintain a hard limit on the number of simultaneously executing jobs.

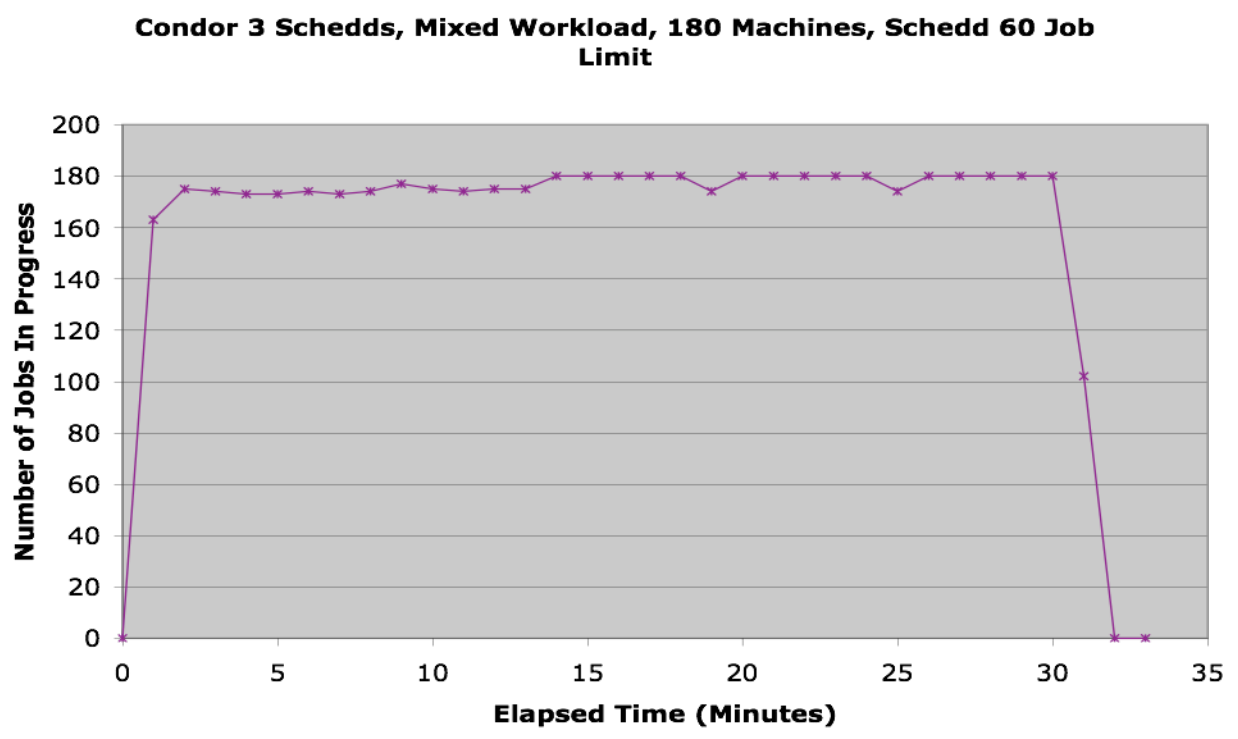

**Figure 16. Condor Mixed Workload, Schedd Limited**

Figure 16 shows the same graph for the same experiment only this time each *schedd* is configured to manage a maximum of 60 simultaneously executing jobs. The throughput under this configuration is close to optimal. In this scenario the *negotiator* allocates each *schedd* one-third of the cluster. With fewer machines to manage, the *schedd*s are able to keep up with the throughput demand resulting in a well-utilized cluster. The drawback to this approach is that there is now an arbitrary limit placed on the capacity of each *schedd*. Under this configuration, for example, a user wishing to run 180 one-hour jobs must also spread these jobs evenly across all three *schedd*s in order to access the full capacity of the cluster. Submitting to only one *schedd* will arbitrarily limit the user to accessing only 60 machines despite the fact that, for the workload of one-hour jobs, a single *schedd* has the capacity to keep up with the throughput demand. We ran several experiments to verify this phenomenon but to save space the resulting graphs are omitted. The key insight is that distributing the workload across multiple *schedd*s does offer an opportunity for coping with mixed workloads, but only to the extent that job submission patterns are sufficiently homogenous.

## 6. SUMMARY AND FUTURE RESEARCH

In this paper we presented an overview of the CondorJ2 cluster management system. The extent to which CondorJ2 leverages off-the-shelf RDBMS and J2EE Application Server technology differentiates it from other batch-computing systems and allows it to treat cluster management as a data management problem.

We also presented three metrics for characterizing system performance. We presented the results of experiments designed to measure CondorJ2's performance on those metrics. There is some evidence indicating that resource limitations interfered with our ability to experimentally determine scalability limits on our current CondorJ2 deployment. In spite of this, the initial results were encouraging. Finally, to place the CondorJ2 results in context, we presented the results of some experiments designed to measure Condor's performance on the same three metrics.

In the future we plan to conduct additional performance and scalability experiments to better understand how CondorJ2 behaves as it scales up. We may employ simulation-modeling techniques if scalability limits continue to exceed what we can test with available resources. As for functionality, work to add user data-set (i.e., the inputs and outputs of the computational jobs that run on the cluster) management services is in progress. We envision a system that uses k-safety, caching and replication to enable more efficient scheduling while also relieving the user of much of the data management burden. Additionally we would like to provide a set of data provenance services. Users would access these services to answer questions like "What executable and input data generated this particular output data set and which versions of the executable and input(s) were used?" We believe that this type of integrated computation and data management system would be a valuable tool for scientists and researchers.

## 7. REFERENCES

[1] Bulhões, P. T., Byun, C., Castrapel, R. and Hassaine, O; "N1 Grid Engine 6 Features and Capabilities"; SUPerG, Phoenix, AZ; May 2004.

[2] Condor Team, Condor Version 6.7.7 Manual, University of Wisconsin – Madison, 2005.

[3] DeMichiel, Linda G., et. al., "Enterprise JavaBeans Specification, Version 2.1," Sun Microsystems, Santa Clara, CA, November, 2003.

[4] International Business Machines Corporation; Tivoli Workload Scheduler LoadLeveler v3.3.2 Using and Administering; International Business Machines Corporation; Poughkeepsie, NY; April 2006.

[5] Litzkow, M., Livny, M., Mutka, M., "Condor – A Hunter of Idle Workstations", in *Proceedings of the Eighth International Conference of Distributed Computing Systems*, 1988.

[6] Litzkow, M.; "Remote Unix – Turning Idle Workstations into Cycle Servers", *Proceedings of Usenix Summer Conference*; pp 381-384; 1987.

[7] Matena, V., Krishnan, S., DeMichiel, L., Stearns, B., Applying Enterprise Java Beans: Component-Based Development for the J2EE Platform, Boston, Addison-Wesley, 2003.

[8] PBS Professional Web Site, http://www.altair.com/software/pbspro.htm.

[9] Platform LSF HPC Web Site, http://www.platform.com/Products/Platform.LSF.Family/Platform.LSF.HPC/.

[10] Raman, R., Livny, M., Solomon, M., "Matchmaking: Distributed Resource Management for High Throughput Computing", in *Proceedings of the Seventh IEEE International Symposium on High Performance Distributed Computing*, Chicago, IL, 1988.

[11] Shannon, B., "Java 2 Platform Enterprise Edition Specification, v1.4, Sun Microsystems, Inc.; Santa Clara, CA; November 2003.

[12] van Engelen, R. and Gallivan, K., "The gSOAP Toolkit for Web Services and Peer-To-Peer Computing Networks," in *Proceedings of the IEEE CCGrid Conference*, 2002.